\UseRawInputEncoding
\documentclass[%
 reprint,
 amsmath,amssymb,
 aps, pre,
]{revtex4-2}

\usepackage{graphicx}
\usepackage{dcolumn}
\usepackage{bm}

\begin{document}

\preprint{APS/123-QED}

\title{Low complexity model to study scale dependence of phytoplankton dynamics in the tropical Pacific}

\author{Jozef Sk\'akala$^{1,2}$}
\author{Paolo Lazzari$^{3}$}%
\affiliation{%
$^{1}$Plymouth Marine Laboratory, Prospect Place, PL1 3DH Plymouth, United Kingdom, \\
$^{2}$National Centre for Earth Observation, PL1 3DH Plymouth, United Kingdom,
}%

\affiliation{$^{3}$National Institute of Oceanography and Applied Geophysics - OGS, Trieste, 34151, Italy.}


\date{\today}

\begin{abstract}
We demonstrate that a simple model based on reaction-diffusion-advection (RDA) equation forced by realistic surface velocities and nutrients is skilled in reproducing the distributions of the surface phytoplankton chlorophyll in the tropical Pacific. We use the low-complexity RDA model to investigate the scale-relationships in the impact of different drivers (turbulent diffusion, mean and eddy advection, primary productivity) on the phytoplankton chlorophyll concentrations.  We find that in the $1/4^{\circ}$ ($\sim$25km) model, advection has a substantial impact on the rate of primary productivity, whilst the turbulent diffusion term has a fairly negligible impact. Turbulent diffusion has an impact on the phytoplankton variability, with the impact being scale-propagated and amplified by the larger scale surface currents. We investigate the impact of a surface nutrient decline and some changes to mesoscale eddy kinetic energy (climate change projections) on the surface phytoplankton concentrations. The RDA model suggests that unless mesoscale eddies radically change, phytoplankton chlorophyll scales sub-linearly with the nutrients, and it is relatively stable with respect to the nutrient concentrations. Furthermore we explore how a white multiplicative Gaussian noise introduced into the RDA model on its resolution scale propagates across spatial scales through the non-linear model dynamics under different sets of phytoplankton drivers. The unifying message of this work is that the low complexity (e.g. RDA) models can be successfully used to realistically model some specific aspects of marine ecosystem dynamics and by using those models one can explore many questions that would be beyond computational affordability of the higher-complexity ecosystem models.
\end{abstract}

\maketitle




\section{Introduction}

There is no effective scale in ecology \cite{levin1992problem}. New structures and processes appear with every new scale down to the fundamental spatial scale of molecular biology, which is far beyond the reach of our ecosystem models. Our models provide merely an effective description for the ecosystem dynamics (e.g.\cite{heinze2013modeling, ford2018marine}), so that any impact of the model sub-grid processes is either parametrized, or can be represented by a stochastic noise. To be able to correctly describe the impact of sub-grid processes on the model grid scale, it is beneficial to have some understanding of how the ecosystem equations, or ecosystem variables evolve with the spatial, or temporal scale ($\ell$). We will call an approach that provides such understanding a "scaling analysis". Scaling analysis has been largely advanced within the framework of renormalization group (e.g.\cite{wilson1971renormalization, wilson1974renormalization, wilson1975renormalization}) with many fundamental applications across particle physics, solid state physics and complex dynamical systems (e.g.\cite{kardar1986dynamic, sinai1988renormalization, cardy1996scaling, rozenfeld2010small}). Interestingly, renormalization group methods were also applied to Navier-Stokes equations (e.g.\cite{forster1977large}) and reaction-diffusion models (for a recent comprehensive review see \cite{dupuis2020nonperturbative}). The renormalization group turned out to be particularly well suited to describe scale-invariant properties of the examined system and has been widely applied to study critical phenomena and universality (e.g.\cite{ma1975critical, fisher1983scaling, kadanoff1990scaling}).

With the increased ecosystem model resolution, as well as the increased model complexity, more phenomena are included into the ecosystem model. However, a model does not necessarily provide good understanding for all the phenomena it represents. Indeed, understanding  phenomena often requires a specific scale: for example to understand oceanic gyres it is desirable to look at a long-time, spatially large-scale oceanic and atmospheric behavior. Although model that captures ocean mesoscale, or sub-mesoscale dynamics represents also ocean gyres, their behavior remains hidden behind the dominant short-time small-spatial scale eddy signal. Similarly if we managed to run a model on a molecular scale, the eddy behavior would remain hidden behind the thermal fluctuations of the molecules and atoms (and the same type of situation happens if we switched further from the atomic scales to the scales of the current elementary particle theory). Here lies another benefit of the scaling analysis: it provides us with a natural tool to understand diverse phenomena with a wide range of characteristic spatio-temporal scales (e.g. turbulence, geologic processes, climate, financial markets, \cite{kolmogorov1941local, mandelbrot1983fractal, turcotte1997fractals, mandelbrot2013fractals, lovejoy2013weather}), as it simultaneously compares processes across different scales.
 
Apart of improving our ecosystem models and understanding processes, there is also a third potential benefit of scaling analysis: Fine resolution models represent broad range of ecosystem phenomena, but they are computationally expensive. Understanding dynamics across range of scales might also optimize the performance of high resolution models by converting them into multi-scale models (e.g.\cite{grooms2018multiscale}). This means each ``separate'' part of the model dynamics could be represented at the maximum scale where it occurs, eventually leading to substantial reduction in the model computational cost.

 
The main point of this work is to develop a schematic multi-scale understanding for some essential aspects of ecosystem dynamics. This provides  different view-point from the standard ecosystem modelling, where the ecosystem model is understood at some specific (fine-resolution) scale, whilst the larger-scale phenomena always ``emerge'' from the model small-scale complex dynamics (e.g.\cite{levin2005self}). We will show that to get a sufficiently realistic representation of the primary productivity in the tropical Pacific, the high complexity model can be for specific purposes bypassed by a simplified ``toy'' reaction-diffusion-advection (RDA) model. The model adds advection term (as in \cite{birch2007bounding}) to the frequently used reaction-diffusion models based on the Fisher-Kolmogorov-Petrovski-Piskunov equation \cite{fisher1937wave, kolmogorov1937study}. 

We forced the RDA model by the realistic Copernicus Marine Environment Monitoring Service (CMEMS) reanalyses for the surface currents and nutrients, and the model remarkably successfully captures the dynamics of chlorophyll also provided by a CMEMS reanalysis. This is a surprising result: Although simple (often one-dimensional) models based on the RDA equation were often used to address conceptual problems, such as how species survival depends on diffusion rate, advection rate, or on the characteristic patch size occupied by the specie (e.g.\cite{skellam1951random, nelson1998non, petrovskii1999plankton, dahmen2000life, cantrell2001spatial, joo2005population, birch2007bounding, mckiver2009influence, mckiver2011resonant, neufeld2012stirring} and for an overview see \cite{ryabov2008population}), one would assume that a sufficiently realistic marine model must be much more complex than the RDA-like models. The relationship between the ecosystem model skill and the model complexity is non-trivial \cite{fulton2003effect}, however certain minimum amount of model complexity is always assumed; the real world marine biogeochemistry is addressed either by the medium-complexity models \cite{schrum2006development, yool2013medusa, aumont2015pisces}, or by the high complexity models \cite{vichi2007generalized, butenschon2016ersem} that have often tens of state variables and more than hundred parameters. Such assumptions are without any doubt founded, but this paper shows that for some suitably chosen problems of high scientific interest, even the simplest model based on RDA equation is capable to produce surprisingly good approximation to the selected real world ecosystem data. However, the words ``suitably chosen problems'' need to be emphasized here, as we are not replacing the full higher complexity model with its lower complexity surrogate, we are only simulating a specific ecosystem model component (surface phytoplankton dynamics) with a low-complexity model, that is forced by some selected higher complexity ecosystem model variables. This is analogous to the typical situation when higher complexity marine ecosystem models are being forced by a marine physical model, or to the coupled marine physical-biogeochemical model being forced by an atmospheric model.
By applying the low-complexity RDA model one gets the best of both worlds: the advantage of the RDA model is that it is cheap to run, and it depends only on three free parameters, whose impact on chlorophyll distributions can be easily understood, modified and studied across wide range of spatial and temporal scales. In the same time the RDA model appears (within its constrained frame of reference) to be sufficiently realistic for the results of such analyses to be taken seriously. 

The paper is structured as follows: (i) firstly we describe, justify, calibrate and validate the RDA model, (ii) we briefly introduce some basic tools and concepts from the scaling analysis, (iii) we apply the scaling analysis to a range of RDA model simulations with modified nutrient and velocity forcing, as well as modified model parameters, in order to derive the scales of spatial and temporal impact of different drivers on the phytoplankton dynamics, (iv) we analyse how phytoplankton scales as a function of the scaled-down nutrients and mesoscale eddies and (v) we add stochastic perturbations to the RDA model in order to investigate how the model non-linear dynamics propagates the impact of the stochastic noise on phytoplankton concentrations through a range of spatial scales.

\section{Methods}

\subsection{The RDA model}



The growth of biomass starts with the photosynthesis in the autotrophic species, and for marine ecosystems these are the diverse species of phytoplankton. The frequently used proxy quantity for phytoplankton biomass is chlorophyll $a$, with a clear advantage of large volume of ocean-color derived observations available for the ocean surface concentrations of chlorophyll (e.g.\cite{groom2019satellite}). In this work we focus on the chlorophyll dynamics, modelled by a RDA equation expressed as
\begin{widetext}
\begin{equation}\label{E1}
\frac{\partial \rho(t,\vec{x})}{\partial t} = ~-\vec{u}(t,\vec{x})\cdot\nabla{\rho(t,\vec{x})}~+~\kappa_{T}\cdot\nabla^{2}\rho(t,\vec{x})~ + ~P\cdot N(t,\vec{x})\rho(t,\vec{x})~ - ~D\cdot\rho^{2}(t,\vec{x}),    
\end{equation}
\end{widetext}
where $\rho(t,\vec{x})$ represents chlorophyll concentrations, $N(t,\vec{x})$ nutrients, $\vec{u}(t,\vec{x})$ is the current velocity,  $\kappa_{T}$ is the diffusivity parameter, which is at the spatial scales considered in this study dominated by the turbulent diffusivity component, $P$ the net primary productivity (growth) rate and $D$ is the damping (mortality) rate. The turbulent diffusivity parameter ($\kappa_{T}$) describes the integrated effect of sub-grid eddy mixing and determines the rate of small-scale chlorophyll smoothing. The damping rate $D$ integrates phytoplankton loss due to the limitation in resources, mortality, respiration and grazing by higher trophic-level species. $D$ also impacts the degree to which chlorophyll and nutrients are correlated: if substantial phytoplankton concentrations get advected into the low-nutrient areas they die off quickly if the damping rate $D$ is high. Conversely, in the nutrient-rich areas the high rate of damping $D$ will not allow phytoplankton to grow above certain threshold in concentrations, constraining the correlation between $\rho$ and $N$.
Finally, the growth parameter ($P$) describes the rate of photosynthesis. The $P$ parameter determines (for a fixed $D$)  the average levels of chlorophyll ($\langle\rho\rangle$) on the domain. 



\begin{widetext}

\begin{figure}
\hspace*{-1.5cm}
\noindent\includegraphics[width=20cm, height=4cm]{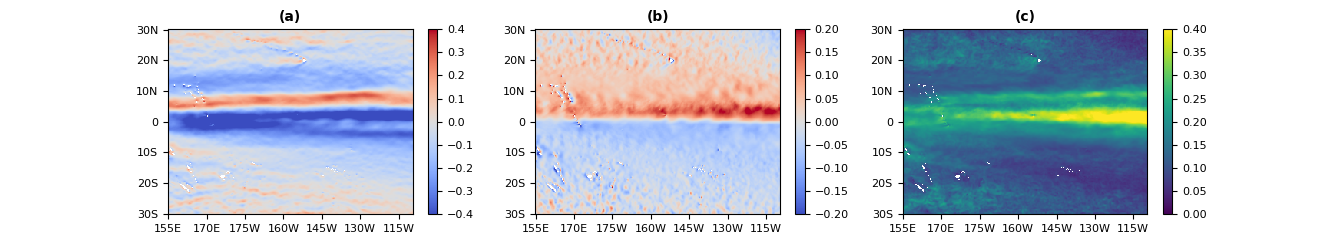}
\caption{The CMEMS horizontal 2017-2018 mean surface current velocity in $m/s$. Left hand panel ($a$) shows the zonal mean current velocity component, the middle panel ($b$) shows the meridional mean current velocity component and the right hand panel ($c$) shows the mean surface eddy speed.}   
\label{fig:one}
\end{figure}

\end{widetext}

For the purpose of this study the RDA model is constrained to a two-dimensional horizontal plane, representing the ocean surface in the Pacific central tropical region (155E - 110W, 30S - 30N, see Fig.\ref{fig:one}). The selected region spans most of tropical Pacific with meridional dimension $\sim$ 6700 km wide and zonal dimension $\sim$ 10600 km long. The RDA model resolution was taken to be $1/4^{\circ}$ ($\sim$25km). The ocean surface current velocity ($\vec{u}(t,\vec{x})$, see Eq.\ref{E1}) and nutrients ($N(t,\vec{x})$) were provided for the RDA model externally; the ocean surface current velocity was taken from the 2017-2018 daily resolution CMEMS reanalysis (GLOBAL\_ANALYSIS\_\-FORECAST\-\_PHY\_001\_024, \emph{http:}\-\emph{//marine.}\emph{copernicus.eu}), which is based on assimilation of satellite sea surface temperature, sea level anomaly, as well as in situ temperature and salinity into $1/12^{\circ}$ ORCA012 model configuration of the Nucleus for European Modelling of the Ocean (NEMO, v3.1, \cite{madec2015nemo}, for details on the reanalysis, see \emph{http://marine.copernicus.eu/} \emph{documents/QUID/}\-\emph{CMEMS-GLO-QUID-001-024.pdf}). To represent the surface currents $\vec{u}$ on the $1/4^{\circ}$ RDA model grid we upscaled the CMEMS data from their original ($1/12^{\circ}$) scale of resolution. In the Fig.\ref{fig:one} we show the 2017-2018 mean values of the surface current velocity vector components and also the mean surface eddy speed. The nutrients $N(t,\vec{x})$ have been estimated as a sum of nitrate and phosphate using the outputs of 2017-2018 CMEMS hindcast based on $1/4^{\circ}$ resolution NEMO coupled with the biogeochemical model Pelagic Interactions Scheme for Carbon and Ecosystem Studies (PISCES, GLOBAL\-\_REANALYSIS\_\-BIO\_001\_029, \emph{http://marine.copernicus.eu/}). No assimilation was used in the biogeochemical run. Phosphate and nitrate were the only nutrient data available with the desired resolution, however taking the sum of nitrate and phosphate is only one of multiple seemingly equivalent choices of how to represent the nutrients. Since phosphate and nitrate concentrations are shaped by similar drivers, the two nutrients have been found to be reasonably highly Pearson correlated ($R=0.78$). The correlation between nitrate and phosphate suggests that different choices on how to combine them into a single nutrient function will yield similar results. We explicitly tried some other options such as the square root of the product from nitrate and phosphate and we have found (not shown here) that the results were indeed qualitatively similar to the choice presented in this study. However, for a specific study on
nutrient regulations, at the same computational cost, it could be possible to investigate
other, more realistic physiological formulations for nutrient co-limitation, e.g. following the Liebig rule. 

The RDA model simulated chlorophyll for the 2017-2018 period, taking the chlorophyll initial value conditions (for 01/01/2017) and open sea boundary conditions from the same CMEMS product than the nutrients. We tested the sensitivity of the RDA model to the initial value and boundary conditions, by replacing the CMEMS chlorophyll data with a Gaussian white noise ($\pm 30\%$ variance) around the $0.1$ $mg/m^{3}$ mean. The tests (not shown here) have demonstrated that on the timescale of $\gtrsim$ 80 days the model is insensitive to the used initial value data. Furthermore, the tests have shown that the impact of boundary conditions on the chlorophyll distributions is negligible.



Tropical Pacific is a region responsible for 20\% of world marine productivity \cite{stanley2010net} and it is an important source of $\hbox{CO}_{2}$ emissions to the atmosphere \cite{behrenfeld2006controls, stanley2010net}. Large parts of the region, such as the eastern equatorial Pacific, are characteristic of high-nutrient (nitrate and phosphate) concentrations due to the equatorial upwelling, but comparably low chlorophyll concentrations (the so-called ``high chlorophyll low-nutrient regions'' \cite{chisholm1991controls, dugdale1991low, eppley1992nitrate}). The comparably low primary productivity around the equator is often understood to be caused by the limited resources of iron \cite{behrenfeld2006controls, johnson2010effect, park2018ocean}, although the elevated levels of grazing also may play a role \cite{dugdale1991low, frost1991role}. In the oligotrophic regions further away from the equator (higher latitude than $10^{\circ}$) the conditions are very different and phytoplankton is mostly nutrient-limited \cite{behrenfeld2006controls}.

There are several reasons why tropical Pacific is an optimal choice for our experiment: 

i) It is an open ocean region with little impact of bathymetry on the ecosystem dynamics. 

ii) The 1-st baroclinic Rossby radius is in the tropical Pacific on the scale of 100-s of km \cite{chelton1998geographical} and the eddy scales can get close to $\sim$ 500 km \cite{chelton2011global} (see also Fig.\ref{fig:two}), so the $1/4^{\circ}$ model resolution allows us to see a wide range of interesting scales for the chlorophyll dynamics \cite{hallberg2013using}. 

iii) Phytoplankton dynamics can be fairly complex and have strong seasonal signatures (e.g. spring blooms) due to seasonal variability in the upper ocean mixing and sunlight that drives photosynthesis. However, the seasonal cycles in the tropical Pacific are weak, and phytoplankton production is primarily regulated by the available nutrients (Fig.\ref{fig:three}), with additional impact of advection by the surface currents. Fig.\ref{fig:three} shows that in the tropical Pacific nutrients and chlorophyll are strongly correlated (Pearson correlation, R=0.77), with nutrient spatial geography playing an essential role for the phytoplankton distributions. In such case one can represent the biological productivity as a simple function of the nutrient concentrations, as is done in the RDA model (Eq.\ref{E1}).   

iv) The RDA model used in this study is a single-equation model with externally supplied nutrients. One could argue that the RDA model needs adding a similar dynamical equation for the nutrients, as has been done many times in the literature (e.g.\cite{ryabov2008population}). In the tropical Pacific the nutrient sinks and sources depend largely on the vertical mixing (e.g. equatorial upwelling) and sufficiently near the coastline could reflect other forcing fields, such as the river discharge. In such case it becomes difficult to implement a two-equation nutrient-chlorophyll model without substantially increasing the model complexity. However, we argue that for the purpose of this study the single equation model (Eq.\ref{E1}) is in the tropical Pacific a reasonable approximation to the phytoplankton chlorophyll dynamics. There are two issues here that need to be raised: Firstly, within this study we will explore the impact of the modified CMEMS data for the surface currents ($\vec{u}$) and the turbulent diffusion parameter ($\kappa_{T}$) on the phytoplankton chlorophyll concentrations ($\rho$). The changed surface advection and diffusion can potentially change the nutrient concentrations ($N$) relative to their externally supplied CMEMS values. Secondly, the phytoplankton concentrations change as a function of the modified advection and the changes to the nutrient uptake by the changed phytoplankton ($\rho$) could be another source that modifies the nutrients relative to their supplied CMEMS values. There are, however, two arguments why we could reasonably neglect those changes to the supplied nutrients and still use the CMEMS product: i) The nutrient distributions are much more geographically stable than the chlorophyll (Fig.\ref{fig:three} and Fig.\ref{fig:four}), by which we mean that the nutrient anomalies are relatively small when compared to the nutrient spatial geography estimated from the 2017-2018 mean values (Fig.\ref{fig:three}). The nutrient geographic sinks and sources, which largely correspond to the upwelling and downwelling zones, then consequently play a key role in the representation of the nutrient distributions, with other drivers (such as eddy mixing, or time-fluctuations in the uptake by phytoplankton) playing mostly a secondary role. Moreover, this study will explicitly demonstrate that it makes little difference to the simulated chlorophyll, whether we force the RDA model with a time-changing, or 2017-2018 time-averaged nutrient distributions. ii) A substantial change to the CMEMS phytoplankton chlorophyll concentrations might indeed introduce some changes to the CMEMS nutrients through the uptake. However, the concern of this study are not the changes to the nutrients, but the impact of those nutrient changes on the phytoplankton distributions. Although the single-equation RDA model does not represent the changes to the nutrients, the quadratic damping term in the RDA model (the $D$-term in Eq.\ref{E1}) effectively integrates into the phytoplankton dynamics the impact of the resource limitation due to nutrient uptake.

Although iron is an important limiting factor in some areas of tropical Pacific \cite{dugdale1991low}, the daily products for iron distributions were unavailable and could not be used as part of the RDA model forcing. The limitation by iron was, similarly to the nutrient uptake, included into the RDA model only implicitly as part of the quadratic damping term. The RDA model assumes that any damping effect included in the quadratic term is proportional to the chlorophyll concentration. This can be easily justified for the rate of phytoplankton mortality, nutrient limitation, or for phytoplankton grazers (their density is expected to be proportional to phytoplankton density), and to some degree it can be justified also for the iron limitation, as the chlorophyll concentrations are highest in the iron-limiting equatorial upwelling region (Fig.\ref{fig:three}). However, we acknowledge that representing the iron limitation only implicitly is definitely a shortcoming of the RDA model.  

\subsection{Some analytical results about the RDA model solutions}

In this section we briefly outline some analytical properties of the RDA model, which will be later used to better understand the results of the study. Since advection and turbulent diffusion do not change the spatially averaged chlorophyll concentration $\langle\rho\rangle$, the RDA model (Eq.\ref{E1}) has a simple stochastic steady state ($\partial\langle\rho\rangle/\partial t = 0$) solution: 
\begin{equation}\label{E1.1}
\langle N'(\vec{x})\rho'(\vec{x})\rangle = \langle \rho'^{2}(\vec{x})\rangle,    
\end{equation}
where $N'=P.N$ and $\rho'=D.\rho$. (For a region with boundaries, we assume in Eq.\ref{E1.1} also constant Dirichlet boundary conditions.)
Applying
\[\langle N'.\rho'\rangle = \langle N' \rangle .\langle \rho' \rangle + Cov(N',\rho'),\]
where ``Cov'' is covariance, or consequently
\[\langle \rho'^{2}\rangle = \langle \rho' \rangle^{2} + Var(\rho'),\]
where ``Var'' stands for variance,
one can transform the stochastic steady state solution (Eq.\ref{E1.1}) into a quadradic polynomial equation for $\langle\rho\rangle$:
\begin{equation}\label{quadratic}
\langle\rho'\rangle^{2}-\langle N' \rangle \langle\rho'\rangle + Var(\rho') - Cov(N',\rho') = 0.
\end{equation}
By solving Eq.\ref{quadratic} we obtain a relationship between the average chlorophyll and the nutrient concentrations as:
\begin{widetext}
\begin{equation}\label{E1.2}
\langle\rho'\rangle ~=~ \frac{\langle N'\rangle}{2}~ \pm ~\sqrt{\left(\frac{\langle N'\rangle}{2}\right)^{2} + Var(\rho')\left(\sqrt{\frac{Var(N')}{Var(\rho')}}\cdot R(\rho',N')-1\right)},   
\end{equation}
\end{widetext}
with $R$ being the Pearson correlation coefficient. A simpler relationship between $\langle\rho'\rangle$ and $\langle N'\rangle$ can be derived, if we assume that the standard deviation of both $\rho'$ and $N'$ is directly proportional to their mean values: $\sqrt{Var(\rho')}=c_{\rho'}.\langle\rho'\rangle$ and $\sqrt{Var(N')}=c_{N'}.\langle N'\rangle$, which is reminiscent of Taylor's law \cite{taylor1982comparative} and it is to some degree supported by ecological data \cite{kilpatrick2003species}. Then Eq.\ref{quadratic} leads directly to a linear relationship:
\begin{equation}\label{slope}
\langle\rho'\rangle = \frac{1+R(\rho',N').c_{\rho'}.c_{N'}}{1+c_{\rho'}^{2}}\cdot\langle N'\rangle.    
\end{equation}
If we lower advection, chlorophyll becomes highly correlated with nutrients and the Pearson correlation $R(\rho',N')$ in Eq.\ref{slope} approaches $R(\rho',N')=1$, whereas with the high levels of mixing $\rho'$ and $N'$ decorrelate ($R(\rho',N')\rightarrow 0$). Eq.\ref{slope} then implies that increasing advection, whilst maintaining the same $\langle N'\rangle$, lowers the mean chlorophyll concentrations. 

If there is neither advection, nor turbulent diffusion ($\vec{u}=\kappa_{T}=0$), and $N'$ does not depend on time, Eq.\ref{E1} has the following exact solutions:
\begin{equation}\label{E1.35}
\rho'(t,\vec{x}) = \frac{N'(\vec{x})}{1 + \rho_{o}\cdot\exp\{-N'(\vec{x})\cdot t\}},
\end{equation}
which converge for $\rho'>0$ to a steady state attractor: 
\begin{equation}\label{E1.4}
\rho'(\vec{x}) =N'(\vec{x}),
\end{equation}
whilst for $\rho'<0$ the solutions runaway to $-\infty$. The solutions from Eq.\ref{E1.4} approach the steady state attractor (Eq.\ref{E1.35}) as:
\begin{equation}\label{E1.8}
\Delta (t,\vec{x}) \simeq \exp\{-N'(\vec{x})\cdot t\},    
\end{equation}
where $\Delta$ is the distance measured on the real line between the approaching solution and the attractor. Eq.\ref{E1.8} means that the higher nutrient concentration, the faster the chlorophyll distributions converge to the steady state solution from Eq.\ref{E1.4}. 

For the exact steady state solution (Eq.\ref{E1.4}) chlorophyll is maximally correlated with nutrients, $R(\rho', N')=1$. A simple consistency check shows that for $R(\rho', N')=1$, Eq.\ref{E1.2} is solved by the averaged form of the linear relationship in Eq.\ref{E1.4}:
\begin{equation}\label{E1.5}
\langle\rho'\rangle =\langle N'\rangle.
\end{equation}
together with
\begin{equation}\label{E1.51}
Var(\rho') = Var(N').
\end{equation}
Eq.\ref{E1.4} and Eq.\ref{E1.5}-\ref{E1.51} imply that if the first two statistical moments of $\rho'$ and $N'$ are equal, then the steady state $\rho'$ and $N'$ are maximally correlated ($R(\rho',N')=1$). With the increased advection $N'$ and $\rho'$ decorrelate, and 
in the limit of $R(\rho',N')=0$, one obtains
\begin{equation}\label{E1.7}
\langle\rho'\rangle ~=~ \frac{\langle N'\rangle}{2} \pm \sqrt{\left(\frac{\langle N'\rangle}{2}\right)^{2} - Var(\rho')}.     
\end{equation}

\vspace{0.5cm}


\subsection{The scaling analysis}
In this work, we borrow insights from the long history of the studies on turbulence and multifractals (\cite{kolmogorov1941local, novikovea1964intermittency, yaglom1966influence, obukhov1968structure, mandelbrot1974intermittent, frisch1978simple, schertzer1987physical, schertzer2011multifractals, lovejoy2013weather}), and use a simple measure for the scale dependence of the system variables ($\Delta_{\ell} \rho$) as: 
\begin{equation}\label{E2}
\Delta_{\ell}\rho = \langle |\rho(x+\ell) - \rho(x)|\rangle.  
\end{equation}
Here $\Delta_{\ell} \rho$ represents a (scale-dependent) magnitude of spatial and temporal variability of $\rho$, $x$ is the spatial, or temporal variable (spatial vector for spatial variability, or time for temporal variability), $\ell$ is the scale of interest and the averaging in Eq.\ref{E2} runs through the relevant spatial domain, or the time interval. $\Delta_{\ell}\rho$ corresponds to the first statistical moment of what is in the multifractal literature often called ``increments'' (e.g.\cite{schertzer2011multifractals}). 

$\Delta_{\ell} \rho$  has the advantage of being methodologically simple and has been many times proven fruitful in the literature (e.g.\cite{schertzer1987physical, schertzer1988multifractal, schmitt1996multifractal, seuront1996multifractal,  lovejoy2001direct, lovejoy2001universal, gagnon2006multifractal, de2011multifractal, lovejoy2013weather}): For the scale-invariant systems the scaling of $\Delta_{\ell}\rho$ follows a power law and it has been found that its power law exponent ($H$) is often an important indicator of the system dynamics, e.g. much research has been carried out to identify the power law exponent for tracers passively advected by a turbulent flow \cite{corrsin1951spectrum, obukhov1968structure, schmitt1996multifractal, de2010passive}. For the intermittent turbulence, the exact value of the tracer power law exponent $H$ depends on some non-trivial assumptions \cite{de2011multifractal}, but the tracers scale sub-linearly, with the phytoplankton power law exponents often in the H=0.33-0.45 range \cite{seuront1996multifractal, de2011multifractal} ($H=0.33$ is the value for the passive tracer in the 3D homogeneous turbulence \cite{corrsin1951spectrum, obukhov1968structure}).

In the recent work of \cite{skakala2019sst} it has been shown that the scaling described by Eq.\ref{E2} is frequently a piece-wise power law with the scaling transition between different power laws corresponding to a transition between different dynamical regimes (see also \cite{seuront1996multifractal, lovejoy2001universal}).
The power law exponents correspond to the scaling slope of $\Delta_{\ell}\rho$ (we use $\tilde\Delta_{\ell}\rho$ notation), which can be analysed by normalizing the $\Delta_{\ell}\rho$ value as 
\begin{equation}\label{scslope}
\tilde\Delta_{\ell}\rho = \Delta_{\ell}\rho/\Delta_{L}\rho,
\end{equation}    
where $L$ is some maximum spatial, or temporal scale of interest \cite{skakala2019sst}. $\tilde\Delta_{\ell}\rho$ can be then used as a simple ``probe'' to test the impact of dynamical drivers (e.g. eddy and mean advection, turbulent diffusion, biological productivity) on the variable of interest (e.g. chlorophyll) across a wide range of spatio-temporal scales.  

\begin{widetext}

\begin{figure}
\hspace*{-1cm}
\noindent\includegraphics[width=15cm, height=8cm]{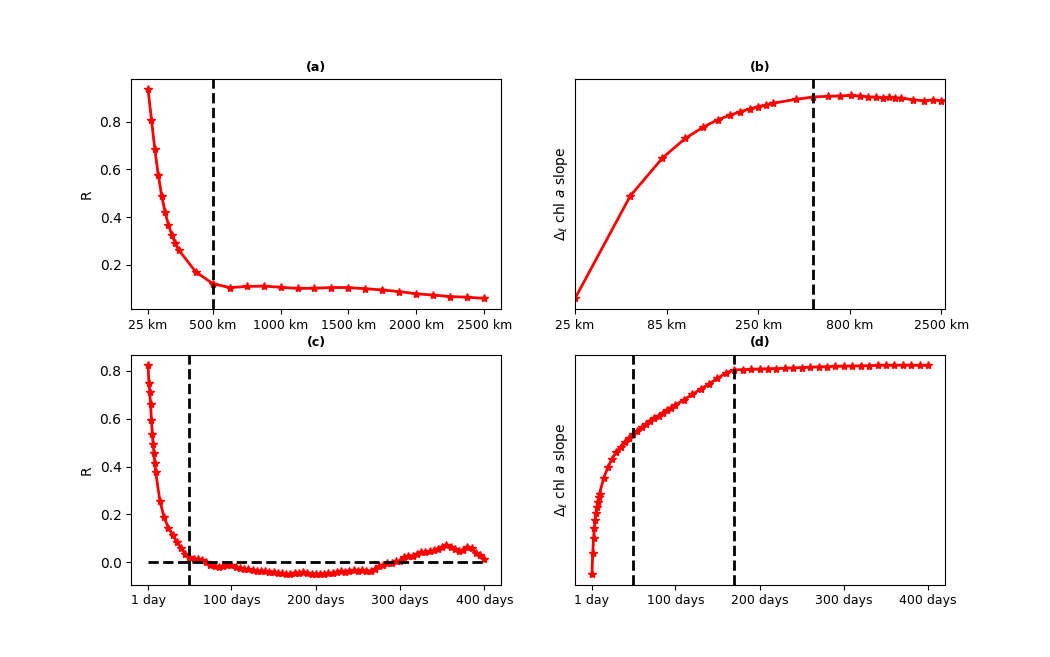}
\caption{The spatial and temporal eddy scales. The left hand side panels ($a$,$c$) show the Pearson correlation ($R$, y-axis) in the eddy surface velocity as a function of spatial ($a$) and temporal ($c$) scale (x-axis), the right hand side panels ($b$,$d$) show the chlorophyll magnitude of log-spatial ($b$) and temporal ($d$) variability ($\Delta_{\ell}\rho$, Eq.\ref{E2}, y-axis) as a function of temporal scale $\ell$ ($d$), or spatial log-scale $\log(\ell)$ ($b$, x-axis). The chlorophyll from the panels $b$ and $d$ was a RDA model output with $\kappa_{T}$=$N$=$0$ and with $\vec{u}$ represented only by the eddy field (the eddy $\vec{u}$ was estimated by subtracting the 2017-2018 mean CMEMS currents from the CMEMS daily output). Since the panels $b$ and $d$ focus only on the scaling slope of $\Delta_{\ell}\rho$, the values of $\Delta_{\ell}\rho$ are not shown. The panel $b$ shows both spatial $\Delta_{\ell}\rho$ and $\ell$ on a log-scale, and it is expected that extending the plot beneath the 25 km scale would yield a power law relationship (straight line on a log-log scale) with an exponent of a passive turbulent tracer. Both analyses ($a$,$c$ and $b$,$d$) point consistently to the maximum eddy spatial scale of 500 km and the maximum time scale of 50 days (this can however be much shorter than eddy life-time, as eddies move). The spatial large-scale correlation ($R\sim 0.1$) that can be seen in the panel a has been found (not shown here) to be caused by a meridional cross-correlation across the equator due to seasonal variations in the currents. Similarly, the $\Delta_{\ell}\rho$ scaling slope within the intermediate time scale between 50-150 days in the panel $d$ has been found (not shown here) to correspond to the seasonal variability in the currents.}   
\label{fig:two}
\end{figure}

\begin{figure}
\hspace*{-2cm}
\noindent\includegraphics[width=20cm, height=9cm]{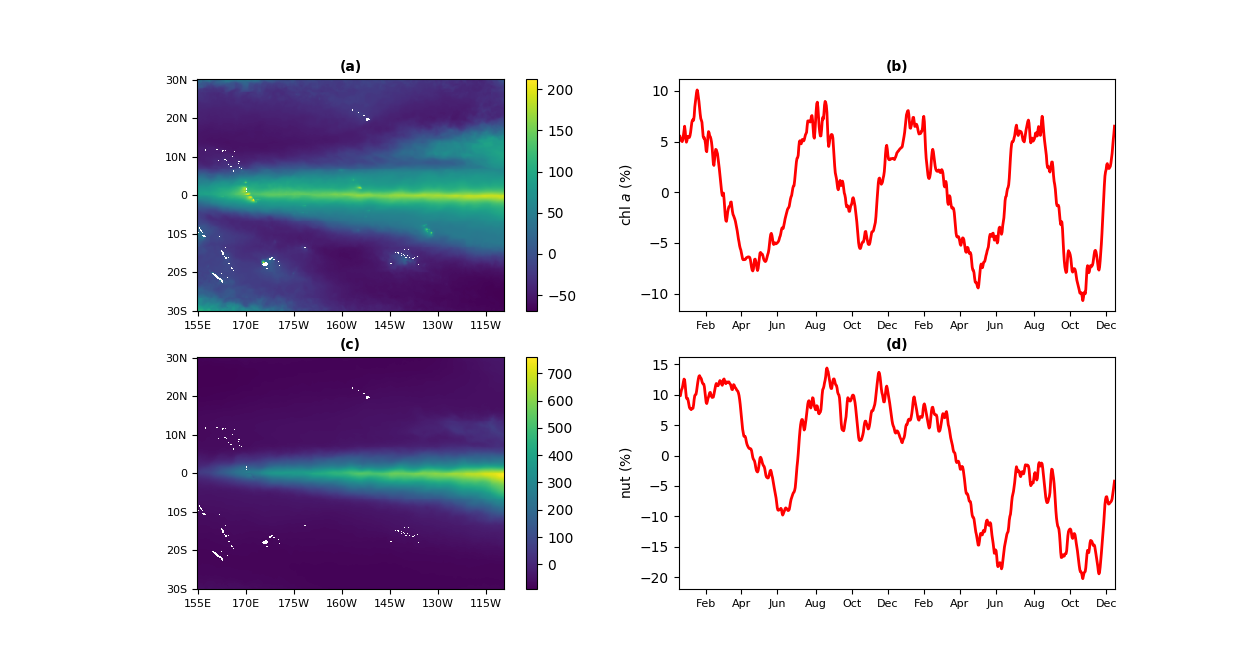}
\caption{The CMEMS 2017-2018 mean surface concentrations for the chlorophyll $a$ (panel $a$) and nutrients (panel $c$), displayed in percentage (\%) deviation from the 2017-2018 mean of the whole spatial domain. The panels $b$ and $d$ show the 2017-2018 time series for the spatial mean of surface chlorophyll ($b$) and nutrients ($d$), displayed in percentage (\%) deviation from the 2017-2018 mean of the whole spatial domain. It is shown that chlorophyll has a modest bi-annual periodicity (panel $b$), which is driven by the seasonal solar cycle (since the region is meridionally symmetric across the equator, the solar seasonal cycle here is bi-annual). 
}
\label{fig:three}
\end{figure}

\begin{figure}
\hspace*{-1cm}
\noindent\includegraphics[width=14cm, height=10cm]{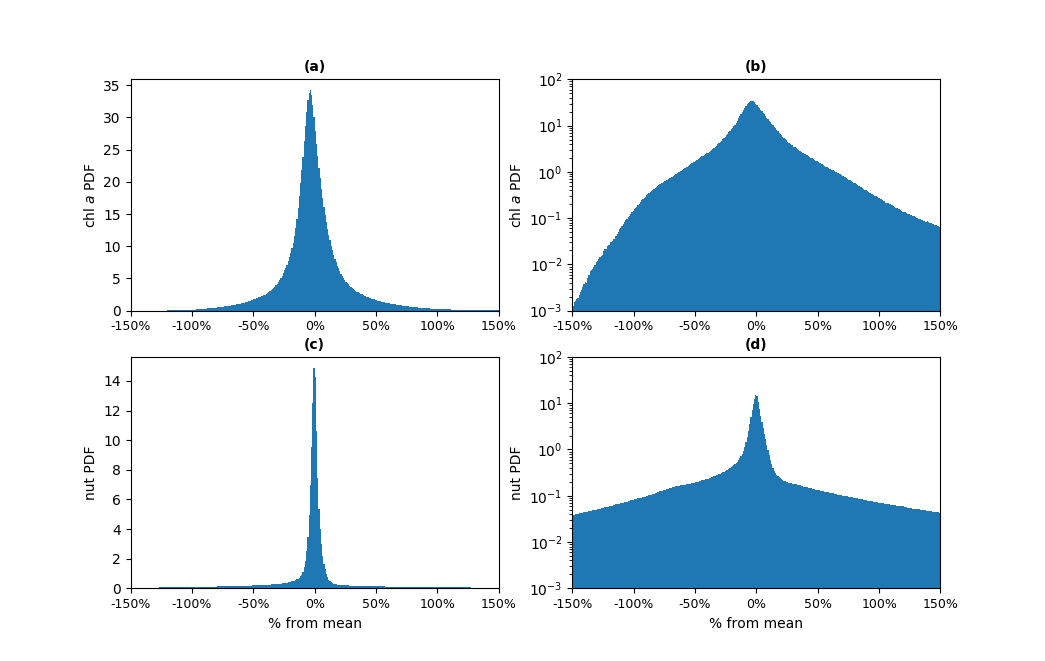}
\caption{The distributions expressed by the Probability Density Function (PDF) for the chlorophyll (panels $a$-$b$) and the nutrient (panels $c$-$d$) anomalies calculated relative to the 2017-2018 mean concentrations. The values on the x-axis are scaled (in \%) relative to the 2017-2018 spatio-temporal mean. The right hand panels ($b$,$d$) show the PDF on a log-scale to emphasize the structure of the tails. The plots show that the relative spread of chlorophyll distribution is substantially larger than the relative spread of nutrients.}   
\label{fig:four}
\end{figure}

\end{widetext}

\section{Validation of the RDA model}

\begin{widetext}

\begin{figure}
\noindent\includegraphics[width=17cm, height=10cm]{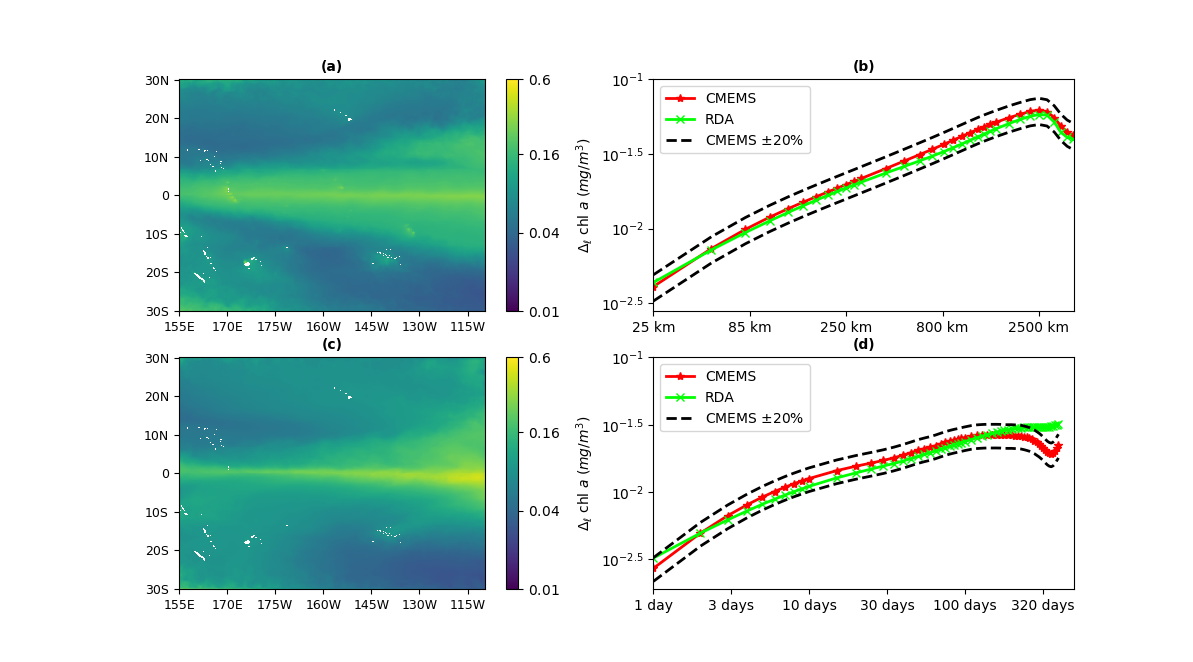}
\caption{The panels compare the CMEMS 2018 mean surface chlorophyll (panel $a$) with the RDA model 2018 mean surface chlorophyll (panel $c$), as well as the chlorophyll magnitude of spatial (panel $b$) and temporal (panel $d$) variability (all in $mg/m^{3}$) across a range of spatial and temporal scales ($\Delta_{\ell}\rho$ calculated as an appropriate average of the 2018 daily data). The panels show that the RDA model is skilled in reproducing both CMEMS chlorophyll spatial distributions (panels $a$ and $c$) and the magnitude of variability (panels $b$ and $d$), except for the magnitude of temporal variability around the half year-to-annual scale. This can be easily explained: the RDA model does not include the time variability in the solar input and hence does not reproduce adequately the bi-annual periodicity of the CMEMS data. The chlorophyll magnitude of spatial variability over 2500 km starts decreasing (panel $b$), since the chlorophyll distributions have a meridional symmetry across the equator. Similarly, as mentioned before, the local minimum of the CMEMS chlorophyll magnitude of temporal variability at the annual scale (panel $d$) is due to the annual cycle (annual cycle seems more pronounced than the bi-annual cycle).}   
\label{fig:five}
\end{figure}

\end{widetext}

An ensemble of RDA model simulations was run until the optimal set of $P$, $D$, $\kappa_{T}$ (Eq.\ref{E1}) values was determined to be:  $\kappa_{T}=300 m^{2}s^{-1}$ (which agrees very well with the values published in the literature \cite{okubo1971oceanic, gargett1984vertical, ledwell1998mixing, zhurbas2004drifter, stewart2008introduction}), $P=7.10^{-8} m^{3}mmol^{-1}s^{-1}$, $D=1.2.10^{-7} m^{3}mg^{-1}s^{-1}$. The set of optimal parameter values was chosen based on the match-ups between the RDA model and the CMEMS data using three metrics shown in Fig.\ref{fig:five}: i) the 2-year mean spatial distribution of chlorophyll,  ii) the magnitude of spatial and temporal variability $\Delta_{\ell}\rho$ (Eq.\ref{E2}) across 25-2500 km and 1 day - 1 year range of scales. The first metric (i, Fig.\ref{fig:five}:$a$,$c$) measures the RDA model skill to estimate the average chlorophyll concentrations and to represent the dominant chlorophyll patterns. Since the spatial chlorophyll patterns dominate over the temporal chlorophyll patterns (Fig.\ref{fig:three}:$a$-$b$) the metric entirely focuses on the chlorophyll spatial distributions. The two remaining metrics (ii, Fig.\ref{fig:five}:$b$,$d$) measure how well the RDA model reproduces the CMEMS magnitude of chlorophyll spatial and temporal variability. The magnitude of chlorophyll spatial and temporal variability will be used to identify the impact of drivers on the chlorophyll concentrations across a wide range of spatial and temporal scales. Since the impact analysis for the chlorophyll drivers relies fully on the RDA model, it is essential that the RDA model reproduces realistically the scaling of the magnitude of chlorophyll spatial and temporal variability. 

All the three metrics in Fig.\ref{fig:five} show that the RDA model is skilled in representing the CMEMS chlorophyll data, i.e. the magnitude of spatial and temporal variablity match on most scales within 10\% and on all scales within 20\%, with the exception of the magnitude of temporal variability on the annual scale. The sudden drop in CMEMS data temporal variability on the annual scale is due to the bi-annual periodicity in the chlorophyll distributions (see Fig.\ref{fig:three}) driven by the bi-annual seasonality pattern in the solar radiation (at the equator the seasonal pattern has bi-annual periodicity, because the seasons in the Southern and Northern Hemisphere have identical impact on the equator). Since the RDA model does not represent the solar cycle, it is understandable that it fails to capture the bi-annual, or annual periodicity in the $\Delta_{\ell}\rho$ of the CMEMS data. 

The RDA model parameters can be characterized by the relative magnitude of three types of drivers: turbulent diffusion, advection and biological activity. The Damk\"ohler number $Da$ (see \cite{birch2007bounding}) gives the scale ($\ell$) dependent ratio between the biological rate of the process and the advection rate: $Da$ = biological rate / advection rate = $\ell.P.\langle N\rangle/\langle|\vec{u}|\rangle$. We can then easily calculate the scale $\ell_{da}$ where biological rate $\approx$ advection rate as $\ell_{da}\approx 3600 km$. At the scales $\ell << \ell_{da}$ advection dominates biological processes and vice versa. If we interpret ``the much smaller'' as a separation by two orders of magnitude, we conclude that advection is expected to dominate biological processes at the O(10) km scales. Similarly to the Damk\"ohler number, we can introduce P\'eclet number \cite{birch2007bounding}, but in the context of turbulent, rather than molecular diffusivity, as $Pe$ = advection rate / turbulent diffusivity rate = $\ell\langle|\vec{u}|\rangle/\kappa_{T}$. Then for the scale $\ell_{pe}$ where advection rate $\approx$ turbulent diffusivity rate, we obtain $\ell_{pe}\approx 700m$. At the scales $\ell >> \ell_{pe}$ advection dominates over turbulent diffusion and vice versa. The $\ell_{pe}$ scale suggests that advection should be dominant over turbulent diffusion on the scales of O(100) km. The estimates using Damk\"ohler and Peclet numbers are broadly consistent with the results of this study, however we will show that advection can propagate the impact of turbulent diffusion (at $1/4^{\circ}$ resolution) to remarkably large scales.

 

\section{Impact of primary productivity drivers across different scales}

\subsection{Spatial analysis}


What will be the impact on the chlorophyll concentrations if we switch off horizontal advection or turbulent diffusion in the RDA model? We have done multiple experiments with: i) switched off mesoscale eddies, in which case $\vec{u}$ (Eq.\ref{E1}) was taken as mean currents only, estimated from a 2017-2018 average of the CMEMS data (see Fig.\ref{fig:one}), ii) switched off mean currents, in which case the 2017-2018 means were subtracted from the CMEMS data for $\vec{u}$ to estimate the eddy field, iii) no advection at all ($\vec{u}=0$). In each of these cases (i-iii) and also in the case forced by CMEMS data for $\vec{u}$ we ran two separate simulations, with and without turbulent diffusion (turbulent diffusion was removed by setting $\kappa_{T}=0$). For the simulation with switched off mesoscale eddies (i), it is desirable to remove the eddy signatures also from the nutrient ($N$) data. We have compared two simulations with the eddy advection $\vec{u}$: a) one that used the CMEMS product for nutrients ($N$) and b) another simulation, which used for $N$ the 2017-2018 mean CMEMS nutrient concentrations. The two simulations produced very similar results for the chlorophyll (not shown here), e.g. the differences in the magnitude of spatial variability were on all scales $<5\%$. In this paper we show the results for the latter simulation (b), but we will keep in mind that those results are representative of both those simulations.     

\begin{widetext}

\begin{figure}
\noindent\includegraphics[width=10cm, height=8cm]{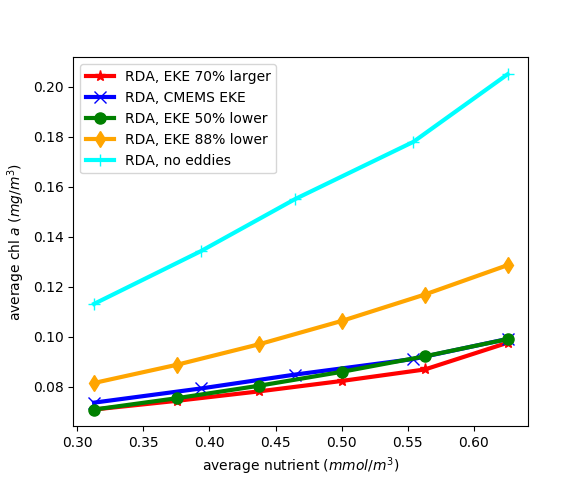}
\caption{The impact of nutrient concentrations on the mean chlorophyll. The values shown in the Figure are the averages through the RDA model spatial domain and the year 2018. It is shown that eddies stabilize the chlorophyll concentration: without eddies the $50\%$ decrease in nutrients leads to almost $50\%$ decrease in chlorophyll, whilst in the presence of CMEMS eddies the $50\%$ decrease in nutrients lowers chlorophyll only by $\sim$ $20\%$.}   
\label{fig:six}
\end{figure}

\end{widetext}

We have observed that advection has substantial impact on the mean chlorophyll values. The levels of chlorophyll increased more than two-fold when mesoscale eddies and diffusion (mostly sub-grid eddy mixing) were removed (Fig.\ref{fig:six}). Furthermore, removing also the mean currents increased chlorophyll concentrations more than three-fold with respect to their original value (not shown). The mean chlorophyll concentration for the zero advection calculated from the RDA model numerical simulation has been found to match remarkably well (on the level of 3\%) with the prediction of the stochastic steady state solution from Eq.\ref{E1.5}-\ref{E1.51}. The limiting impact of the advection (or diffusion) term on the primary productivity is well known in the literature on the RDA-like models (\cite{kierstead1953size, okubo1978advection, okubo1978horizontal, murray1983minimum, holmes1994partial, cantrell2001spatial, speirs2001population, ryabov2008population, okubo2013diffusion}) and can be understood through a simple argument: Take $n_{L}, n_{S}$ and $ch_{N}, ch_{S}$ where $n_{L}, ch_{L}$ are ``large'' nutrient and chlorophyll concentrations, whilst $n_{S}, ch_{S}$ are ``small'' nutrient and chlorophyll concentrations. Since ``large'' is larger than ``small'' we have:
\begin{equation}
(n_{L} - n_{S})\cdot(ch_{L} - ch_{S}) > 0
\end{equation}
implying that
\begin{equation}\label{eq2}
n_{L}ch_{L} + n_{S}ch_{S} > n_{L}ch_{S} + n_{S}ch_{L}.
\end{equation}
Advection (e.g. eddy mixing) brings large chlorophyll concentrations $ch_{L}$ to areas with worse growth conditions (small nutrient concentrations $n_{S}$) and vice versa, the growth term then corresponds to the right side of Eq.\ref{eq2}, whereas if there was no advection the growth term is described by the left side of Eq.\ref{eq2}. This means when there is advection (eddy or mean) the growth term is smaller than if there is no advection. 
However, focusing purely on eddies, their size matters: the eddies that impact primary productivity have to act on a scale with substantial nutrient variability. Otherwise the inequality between the two growth terms in Eq.\ref{eq2} has small impact since $ch_{L}$ and $ch_{S}$ ($n_{L}$ and $n_{S}$) are of comparable size.    
For example the turbulent diffusion term representing eddy mixing beneath the 25 km scale has been found to have very little impact ($\sim 10\%$) on the mean chlorophyll concentration.

By removing (mesoscale \& sub-grid) eddies the magnitude of chlorophyll spatial variability ($\Delta_{\ell}\rho$) increases on all scales roughly fourfold (Fig.\ref{fig:seven}:$b$ and Tab.\ref{table:one}). This is not hugely surprising, since the removal of mesoscale eddies increased primary productivity and doubled the mean chlorophyll concentrations. The increased chlorophyll concentrations then usually imply a higher chlorophyll variability. However, the different scales of eddy impact on the chlorophyll distributions can be estimated from the chlorophyll scaling slopes $\tilde\Delta_{\ell}\rho$ (Eq.\ref{scslope} and see also \cite{skakala2019sst}), rather than directly from the magnitude of chlorophyll spatial variability ($\Delta_{\ell}\rho$, Eq.\ref{E2}).
Eddies should lower variability (steepen the scaling slope) above the characteristic eddy scale (they mix, therefore smooth) and increase variability (flatten the scaling slope) at the range of scales with eddies (due to characteristic eddy patchiness). Assuming that any smoothing effect above the eddy scale goes away at a sufficiently large scale, one can determine the range of scales where the $\tilde\Delta_{\ell}\rho$ differs between the case with and without eddies. Given that above $\gtrsim 500$ km  $\tilde\Delta_{\ell}\rho$ scales with similar slope in both cases (the case with eddies vs the case without eddies), it is natural to assume that the only important impact of eddy patchiness, or eddy mixing, on the chlorophyll variability happens at $\lesssim 500$ km where the removal of eddies steepens the $\Delta_{\ell}\rho$ scaling slope ($10\%$ increase in variability under $\sim 250$ km, due to eddy patchiness, see Fig.\ref{fig:seven} and also Tab.\ref{table:one}). Using the chlorophyll $\tilde\Delta_{\ell}\rho$ from Fig.\ref{fig:seven} and approximating it on the 100-500 km range with a power law, one can estimate the chlorophyll scaling exponent as $H\approx 0.6$, which is somewhat above the previously estimated passive turbulent tracer range at mesoscale \cite{de2011multifractal} and may indicate a mixed active tracer - eddy advection regime. 

It is interesting to analyze the interaction between explicit advection terms
and the sub-grid eddy mixing captured by the turbulent diffusion term (Fig.\ref{fig:eight}, Tab.\ref{table:two}).
Due to mesoscale eddies and large scale currents (``mean'' flows) the smoothing impact of turbulent diffusion spreads to the large spatial scales, i.e. at the resolution $\sim 25$ km scale removing turbulent diffusion more than triples the chlorophyll variability (Tab.\ref{table:two}) and it increases variability by at least 10\% up to 2000 km scale (see Fig.\ref{fig:eight}:$d$). We can then separate out the relative impact of the mesoscale eddies and the mean currents on the large scale smoothing (see Fig.\ref{fig:eight}). With the model advection completely turned off, removing turbulent diffusion increased the magnitude of chlorophyll spatial variability by a maximum 10\% at the resolution scale (Fig.\ref{fig:eight}:$a$, Tab.\ref{table:two}), with a detectable impact on the chlorophyll variability constrained to the $\lesssim$ 70 km scales. By switching on mean currents, but no mesoscale eddies, removing the turbulent diffusion increased the chlorophyll variability by about 100\% at the resolution scale, and the impact of turbulent diffusion on chlorophyll variability lasted up to $\sim 4000$ km (but beneath $10\%$ from 700 km scale, Fig.\ref{fig:eight}:$c$, Tab.\ref{table:two}). Switching on mesoscale eddies but not the mean currents, turbulent diffusion term impacted the chlorophyll variability approximately up to the 600-800 km scale (Fig.\ref{fig:eight}:$b$). Overall the impact of the turbulent diffusion term seemed to be equally amplified by the mesoscale eddies and the mean currents (Fig.\ref{fig:eight}:$b$-$c$, Tab.\ref{table:two}).  

\begin{widetext}

\begin{figure}
\noindent\includegraphics[width=11cm, height=15cm]{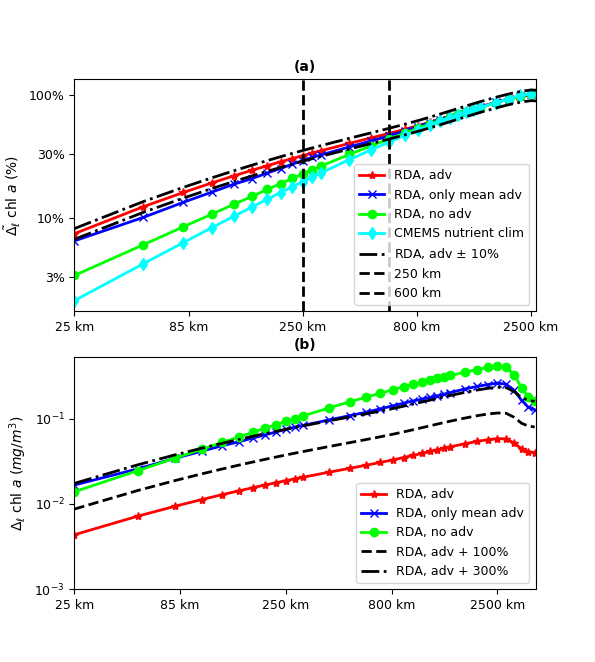}
\caption{The panel $a$ shows the percentage reduction in the magnitude of chlorophyll spatial variability ($\Delta_{\ell}\rho$) when compared to the magnitude of its spatial variability at the scale L=2500km ($\Delta_{L}\rho$), or equivalently it compares the spatial scaling slopes ($\tilde\Delta_{\ell}\rho$) for the different simulations. The panel $b$ shows the absolute values for the magnitude of spatial variability $\Delta_{\ell}\rho$ (in $mg/m^{3}$). The $\Delta_{\ell}\rho$ (panel $b$) and $\tilde\Delta_{\ell}\rho$ (panel $a$) curves represent the 2018 annual averages of the spatial scaling of the daily data. Both x and y axes are on a log-scale. We show the relative (panel $a$) and absolute (panel $b$) chlorophyll magnitude of spatial variability for the different dynamical scenarios of the RDA model: i) model forced by both mean and eddy surface currents (``RDA, adv''), ii) model forced only by the mean currents (``RDA, only mean adv'') and iii) model with all the (eddy and mean) advection removed (``RDA, no adv''). In addition to the chlorophyll variability, the cyan line marked with diamonds in the panel $a$ shows the magnitude of spatial variability for the 2017-2018 averaged nutrient concentrations (``CMEMS, nutrient clim''). The dashed lines parallel to the variability curves mark a 100\% and 300\% increase in the magnitude of spatial variability with respect to the RDA model forced by both eddy and mean advection. The vertical lines show the scales from which the relative scaling remains within $10\%$ from the fully (eddy \& mean advection) forced RDA model. 
}   
\label{fig:seven}
\end{figure}

\begin{figure}
\noindent\includegraphics[width=18cm, height=12cm]{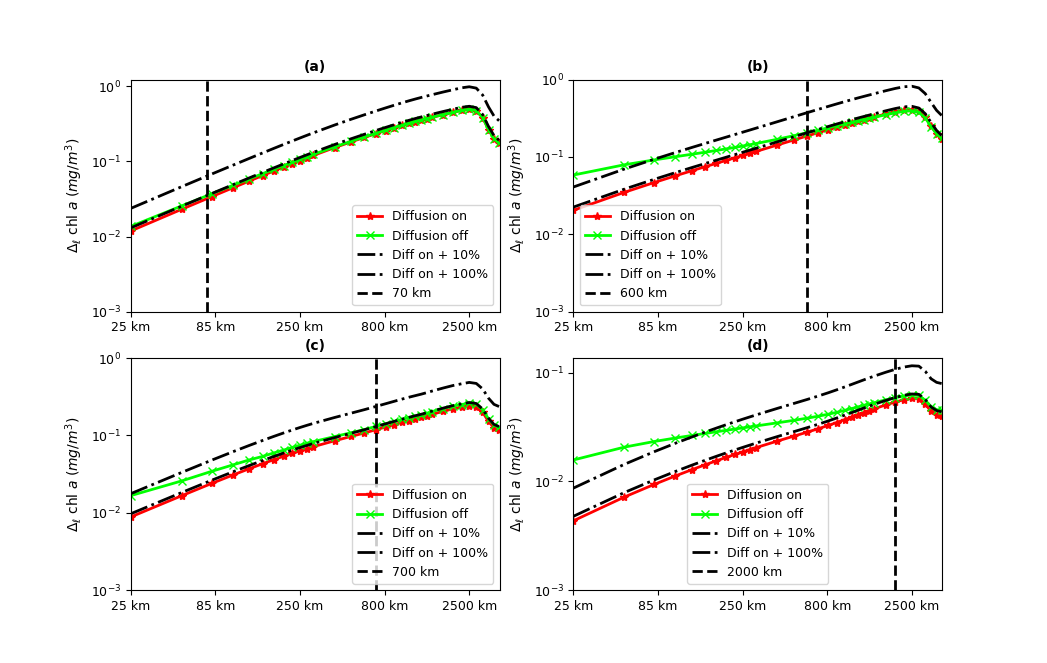}
\caption{The impact of the turbulent diffusion term in the RDA model on the chlorophyll magnitude of spatial variability (2018 averages from daily $\Delta_{\ell}\rho$, all in $mg/m^{3}$) depending on the advection input: no advection (panel $a$), only mesoscale eddy advection (panel $b$), only mean advection (panel $c$), both mean and mesoscale eddy advection (panel $d$). The dashed lines parallel to the $\Delta_{\ell}\rho$ curves mark a 10\% and 100\% increase in the magnitude of spatial variability with respect to the RDA model with the turbulent diffusion. The vertical lines show the scale from which $\Delta_{\ell}\rho$ matches the fully forced model within $10\%$. When there is no advection the turbulent diffusion term has a spatially limited impact up to $\sim$ 70 km scale. With mesoscale eddies and/or mean currents the impact of turbulent diffusion on the chlorophyll magnitude of spatial variability increases 2-4 times at the resolution scale and becomes substantial up to 600-700 km scale with mean currents having non-negligible impact up to the largest scale ($\sim$ 4000 km). The overall impact of mesoscale eddies and mean currents on the turbulent diffusion term is comparable.}   
\label{fig:eight}
\end{figure}
  

\begin{table}[!htbp]
\caption{We show the impact of different drivers on the chlorophyll magnitude of spatial variability ($\Delta_{\ell}\rho$). The Table shows the values displayed in Fig.\ref{fig:seven} and Fig.\ref{fig:eight}. The first column shows the percentage change in the magnitude of spatial variability at 2500 km after we removed a specific driver (turbulent diffusion, eddy and mean advection) from the fully forced RDA run. The numbers in the first column amount to the comparison of the different curves from Fig.\ref{fig:seven}:$b$ at 2500 km, and the purpose of those numbers is to show the overall change to the spatial variability at the regional scale. The second-to-fourth column display the percentage change to the spatial scaling slopes $\tilde\Delta_{\ell}\rho$ (the scaling slopes are understood as a ratio $\Delta_{\ell}\rho/\Delta_{L}\rho$ with L = 2500 km, see Fig.\ref{fig:seven}:$a$) in the situation without a specific driver when compared to the fully forced RDA model. The percentage change is shown for a range of values within three intervals of spatial scales: 25-100 km, 100-500 km and 500-2500 km. The $\uparrow\downarrow$ symbols before the numbers indicate whether the RDA value increases ($\uparrow$), or decreases ($\downarrow$) when the specific driver is removed.}
\centering
\vspace{0.3cm}
\begin{tabular}{c c c c r }
\hline
{\bf removed driver} & $\Delta_{\ell}\rho$ at 2500 km & 25-100 km & 100-500 km & 500-2500 km \\ 
\hline
diffusion & $\uparrow$ 10\% & $\uparrow$ 108-238\% & $\uparrow$ 30-108\% & $\uparrow$ 0-30\% \\ 
diffusion \& eddy & $\uparrow$ 340\% & $\downarrow$ 13-16\% & $\downarrow$ 6-16\% & $\downarrow$ 0-6\% \\
mean advection &  $\uparrow$ 580\% & $\downarrow$ 25-30\% & $\downarrow$ 6-25\% & $\downarrow$ 0-6\% \\
diffusion \& eddy \& mean & $\uparrow$ 600\% & $\downarrow$ 44-54\% & $\downarrow$ 14-44\% & $\downarrow$ 0-14\% \\
\hline
\end{tabular}
\label{table:one}
\end{table}

\begin{table}[!htbp]
\caption{We show how the different drivers (eddy and mean advection) propagate the impact of turbulent diffusion on the chlorophyll magnitude of spatial variability ($\Delta_{\ell}\rho$), as displayed in the Fig.\ref{fig:eight}. The first column shows the percentage change in the magnitude of spatial variability at 2500 km between the runs with and without turbulent diffusion, after we removed a specific driver (turbulent diffusion, eddy and mean advection) from the fully forced RDA run. The numbers in the first column amount to the comparison of the pairs of curves from Fig.\ref{fig:eight}:$a$-$d$ at 2500 km, and the purpose of those numbers is to show the overall change to the spatial variability at the regional scale. The second-to-fourth column display the percentage change to the spatial scaling slopes $\tilde\Delta_{\ell}\rho$ (the scaling slopes are understood as a ratio $\Delta_{\ell}\rho/\Delta_{L}\rho$ with L = 2500 km, see Fig.\ref{fig:seven}:$a$) in the situation with and without turbulent diffusion after a specific driver was removed from the fully forced RDA model. The percentage change is shown for a range of values within three intervals of spatial scales: 25-100 km, 100-500 km and 500-2500 km. The $\uparrow\downarrow$ symbols before the numbers indicate whether the RDA value increases ($\uparrow$), or decreases ($\downarrow$) when the specific driver is removed.}
\centering
\vspace{0.3cm}
\begin{tabular}{c c c c r }
\hline
{\bf removed driver} & $\Delta_{\ell}\rho$ at 2500 km & 25-100 km & 100-500 km & 500-2500 km \\ 
\hline
none & $\uparrow$ 10\% & $\uparrow$ 108-238\% & $\uparrow$ 30-108\% & $\uparrow$ 0-30\% \\ 
eddy & $\uparrow$ 10\% & $\uparrow$ 26-75\% & $\uparrow$ 3-26\% & $\uparrow$ 0-3\% \\
mean advection & 0\% & $\uparrow$ 83-196\% & $\uparrow$ 18-83\% & $\uparrow$ 0-18\% \\
eddy \& mean & 0\% & $\uparrow$ 7-15\% & $\uparrow$ 1-7\% & $\uparrow$ 0-1\% \\
\hline
\end{tabular}
\label{table:two}
\end{table}
\end{widetext}


\subsection{Temporal analysis}

\begin{widetext}

\begin{figure}
\noindent\includegraphics[width=18cm, height=12cm]{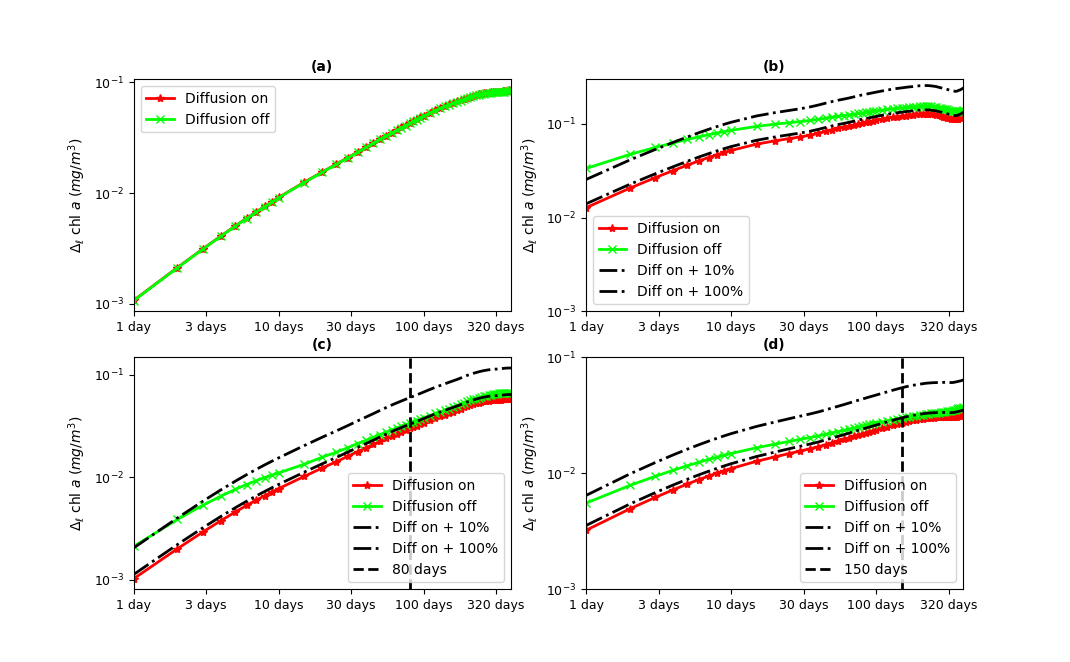}
\vspace*{.5cm}
\caption{
The impact of the turbulent diffusion term in the RDA model on the chlorophyll magnitude of temporal variability (in $mg/m^{3}$) depending on the advection input: no advection (panel $a$), only mesoscale eddy advection (panel $b$), only mean advection (panel $c$), both mean and mesoscale eddy advection (panel $d$). The dashed lines parallel to the variability curves mark a 10\% and 100\% increase in the magnitude of temporal variability with respect to the RDA model with the turbulent diffusion. The vertical lines show the scale from which the variability matches the fully forced model within $10\%$. The combined impact of mesoscale eddies and mean currents on how turbulent diffusion spreads across time-scales is highly non-linear: removing mean currents close to the daily time-scale impacts chlorophyll temporal variability more (in both absolute numbers and proportionally) than removing both mean currents and mesoscale eddies. If there is no advection, turbulent diffusion has no impact on the chlorophyll temporal variability above the daily scale (upper left panel). By including mean currents, but no mesoscale eddies, turbulent diffusion increases the magnitude of chlorophyll temporal variability by $>$ $10\%$ on the full range of scales (1 day - 1 year). By including eddy advection, but no mean currents, turbulent diffusion has $<$ $10\%$ impact on the magnitude of chlorophyll temporal variability above the $\sim$ $80$ day scale, which is broadly consistent with Fig.\ref{fig:six}. Removing both mesoscale eddies and mean currents has $<10\%$ impact on the magnitude of chlorophyll temporal variability above the scale of a half year.}   
\label{fig:nine}
\end{figure}

\end{widetext}

Fig.\ref{fig:nine} shows an exact analogue of Fig.\ref{fig:eight} with the temporal scaling replacing the spatial scaling. It is shown that with no advection, the turbulent diffusion term has a negligible effect on the magnitude of chlorophyll temporal variability above the daily scale (Fig.\ref{fig:nine}:$a$). The impact of advection on the chlorophyll diffusive smoothing (Fig.\ref{fig:nine}:$c$-$d$) appears highly non-linear: The largest effect is observed due to the mean currents and this effect is perhaps surprisingly reduced when also eddies are removed (Fig.\ref{fig:nine}:$d$). However, more broadly the conclusions based on the temporal analysis (Fig.\ref{fig:nine}) are consistent with the spatial analysis (Fig.\ref{fig:eight}). Fig.\ref{fig:nine} confirms that advection substantially increases the impact of turbulent diffusion on the chlorophyll variability on a large range of scales ($>$ $10\%$ for up to the 180 day time-scale). 

\subsection{A relationship between chlorophyll spatial and temporal scales}

\begin{widetext}

\begin{figure}
\noindent\includegraphics[width=12cm, height=18cm]{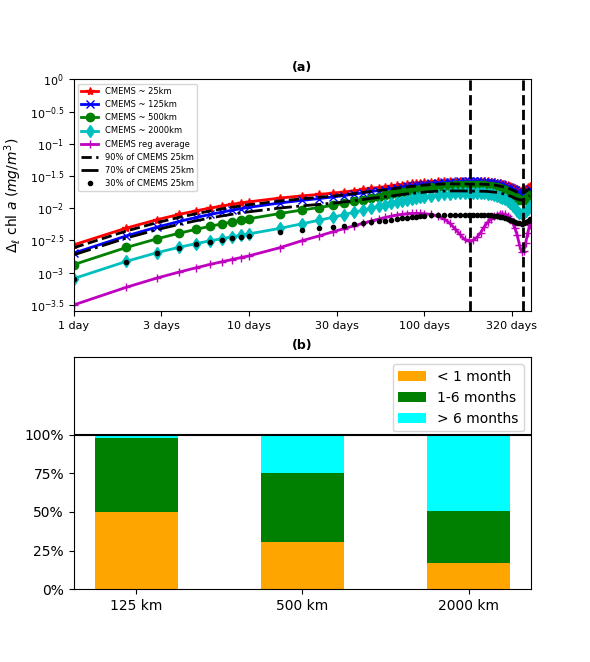}
\caption{The upper panel ($a$) shows the magnitude of temporal variability (in $mg/m^{3}$) for the spatially filtered CMEMS chlorophyll data (moving median filter) at a range of scales: 25 km (original resolution), 125 km, 500 km, 2000 km, and the magnitude of temporal variability of the regional mean value (``CMEMS reg. average''). The two local minima (dashed vertical lines) in the curve for the regional mean value correspond to the bi-annual periodicity in the CMEMS chlorophyll. The scale where the temporal variability of the spatially filtered data meets with the temporal variability of the original CMEMS (25 km) data is the scale where the processes removed by the spatial filtering have no longer impact on the magnitude of chlorophyll temporal variability. The upper panel then provides connection between the spatial and the temporal scales shown in the bottom panel ($b$). The bottom panel demonstrates how the temporal variability of the spatially filtered data (at 125 km, 500 km and 2000 km) splits (in \%) into three categories: $<$ 1 month variability, 1-6 months variability and $>$ 6 months variability. We see that when we remove processes beneath 125 km the $>$ 6 months variability is only $2\%$ of the missing daily variability, whilst in the case of 2000 km spatial filter it grows to $50\%$.}   
\label{fig:ten}
\end{figure}

\end{widetext}

The chlorophyll distributions are influenced by the complex dynamics occurring at wide ranges of spatial and temporal scales. To have a simultaneous understanding of the ecosystem processes across a range of spatio-temporal scales, it is of general interest to find a relationship between the characteristic spatial and temporal scales for the processes driving surface chlorophyll. In this short section we will not distinguish between the specific processes driving chlorophyll, but we will demonstrate (Fig.\ref{fig:ten}) a methodology (developed in \cite{skakala2019sst}) on how to find a relationship between the spatial and the temporal scales for the magnitude of chlorophyll variability. In essence, the relationship is defined by computing the magnitude of temporal variability for a sequence of low pass filtered CMEMS chlorophyll spatial distributions at a range of spatial scales (125 km, 500 km, 2000 km). Spatial filtering removes processes that occur on the sub-filter spatial scales and those processes typically influence the chlorophyll dynamics on some specific range of temporal scales. For example, the processes removed by the spatial filtering may lead to a substantial decrease in the CMEMS chlorophyll daily variability (Fig.\ref{fig:ten}). As one increases the temporal scale, the spatial high-resolution scale processes that were removed by the low pass filter play lesser role in the magnitude of chlorophyll temporal variability ($\Delta_{\ell}\rho$) and the $\Delta_{\ell}\rho$ curves of the spatially filtered and the unfiltered chlorophyll start converging to each other. This means that the difference in the daily variability between the filtered and the unfiltered chlorophyll (we will call it ``Missing Daily Variability of the Filtered Data'' and abbreviate it with MDVFD) is reduced when we increase the temporal scales. The connection between spatial and temporal variability is provided as follows: For each spatial filter (at spatial scale $\ell$) we subdivide the temporal scales into different ranges ($<1$ month, $1-6$ months, $>6$ months) and ask how much was MDVFD reduced at each specific range of temporal scales. Then if MDVFD reduces by $N\%$ at a certain range of temporal scales (e.g 1-6 months) then we say that this specific range of temporal scales contains $N\%$ of MDVFD. It is then clear that as one increases the spatial scale of the low pass filter one removes processes with longer temporal scales and larger fraction of MDVFD will be concentrated at larger temporal scales (e.g above the scale of 6 months). This provides a connection between the spatial scale of the low pass filter and the ranges of temporal scales of MDVFD. The Fig.\ref{fig:ten} shows this spatio-temporal relationship: while the 125 km spatial filter has 50\% of MDVFD on sub-monthly scales and only 2\% of MDVFD on scales larger than half year, the 2000 km spatial filter corresponds to 17\% of MDVFD on the sub-monthly scale and almost 50\% of MDVFD on the scales larger than half year.

\subsection{Impact of nutrients and eddies on chlorophyll}

In the climate change scenarios the upper ocean warms up, leading to increased ocean stratification. The increasingly stratified ocean acts as a barrier to vertical nutrient mixing and lowers the surface nutrient concentrations \cite{lozier2011relationship, capotondi2012enhanced}. Besides nutrients, the increased vertical stratification influences the first baroclinic Rossby radius impacting on the mesoscale eddy kinetic energy (EKE, \cite{saenko2006influence, sueyoshi2009reproducibility}). In this section we will use the RDA model to explore the impact of the changed nutrients and EKE on the surface chlorophyll. Although the RDA model is a major simplification, we believe it might offer at least some qualitative insights into how phytoplankton might respond to some of the environmental changes.  
A form of analytical relationship between the mean chlorophyll and the mean nutrients has been derived for the stochastic steady state of the RDA model in Eq.\ref{E1.2}. 
However, in reality chlorophyll might be far from a stochastic steady state prediction described by Eq.\ref{E1.2} and we have found (not shown here) that the stochastic steady state model does not approximate well the simulations from this study. 

In Fig.\ref{fig:six} we show the spatio-temporal means (for 2018 and the RDA spatial domain) of chlorophyll and nutrients plotted against each other in a series of experiments, where the CMEMS nutrients and EKE (forcing the RDA model) were re-scaled by constant factors, i.e. as $k.N(t,\vec{x})$, where $N$ are the nutrients from the CMEMS model. The constant ($k$) factors were for nutrients taken from the $k\in (0.5,1.4)$ interval and for EKE from the $k\in(0,1.7)$ interval (in case of EKE we rescaled each eddy velocity component with the same factor). 
Fig.\ref{fig:six} demonstrates that changing EKE by $\pm$ 50-70\% has a relatively minor impact on the mean chlorophyll concentrations, whilst lowering EKE more substantially (by $\sim$ 85\% and more) can have a large impact on the mean chlorophyll concentrations. Fig.\ref{fig:six} also demonstrates that under the increased EKE, phytoplankton becomes increasingly insensitive to the changing nutrients: with zero EKE chlorophyll scales almost linearly with nutrients (e.g. 50\% decrease in nutrients amounts to 50\% decrease in chlorophyll concentrations, similar to Eq.\ref{slope}), while with increased EKE the scaling becomes increasingly sub-linear (e.g. for CMEMS EKE 50\% decrease in nutrients there is about 20\% decrease in chlorophyll concentrations). 
This is an interesting result implying that in the increased EKE scenario the phytoplankton concentrations become more stable. In particular, the RDA model suggests (Fig.\ref{fig:six}) that within $\pm 70\%$ of the current EKE levels, a dramatic decline of nutrients has comparably small impact on chlorophyll. 
It is not entirely clear how to interpret this result, neither how seriously it should be taken: we would recommend to take it with a lot of caution, unless it is reconfirmed in more realistic simulations using higher complexity models.

\begin{widetext}

\begin{figure}
\hspace*{-1cm}
\noindent\includegraphics[width=20cm, height=13cm]{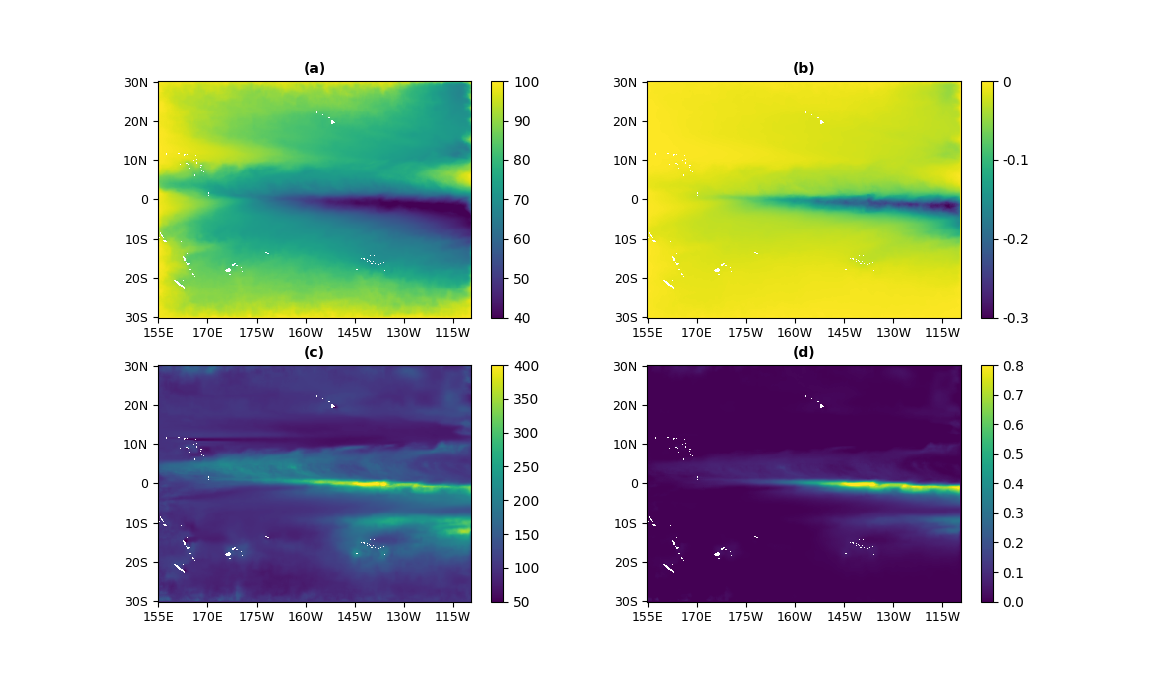}
\caption{Panels $a$-$b$ show the impact of $50\%$ nutrient decrease on the annual 2018 mean chlorophyll concentrations. Similarly to $a$-$b$, the panels $c$-$d$ show the impact of 88\% decrease in the EKE on the mean annual 2018 chlorophyll concentrations. The panels $b$ and $d$ show the absolute change (in $mg/m^{3}$) in chlorophyll concentrations when compared to the simulation forced by the CMEMS nutrients and EKE.  The panels $a$ and $c$ show the same change, but relative (in \%) to the values of the simulation using the CMEMS data.}   
\label{fig:eleven}
\end{figure}

\end{widetext}

It is also interesting to explore how the chlorophyll surface distributions respond to the changes imposed on the nutrients, or EKE. 
Since in tropical Pacific the chlorophyll spatial variability dominates over temporal variability (Fig.\ref{fig:three}) it is useful to understand how the spatial regional patterns of chlorophyll change under the changed chlorophyll mean. In Fig.\ref{fig:eleven} we show the impact of halved nutrient concentrations (Fig.\ref{fig:eleven}:$a$-$b$) and decreased EKE by 88\% (Fig.\ref{fig:eleven}:$c$-$d$) on the chlorophyll annual mean spatial distributions. The changed nutrient concentrations and the eddy velocities were re-scaled versions of the original CMEMS data, where by ``re-scaled'' we mean the original CMEMS distributions multiplied by a spatio-temporally constant factor. The Fig.\ref{fig:eleven} shows that the resulting chlorophyll 2018 mean spatial distributions are far from being the re-scaled versions of the 2018 mean chlorophyll forced by the CMEMS data. In particular Fig.\ref{fig:eleven}:$a$ shows that under the nutrient decline chlorophyll changes by substantially larger proportion in the areas with higher chlorophyll concentrations (eastern tropical Pacific). This indicates that areas with the highest biological activity are also most vulnerable to change.  
It is perhaps surprising that reducing nutrients (Fig.\ref{fig:eleven}:$a$-$b$) has proportionally largest impact on chlorophyll in the chlorophyll-rich areas, since the same areas have correspondingly highest eddy activity (Fig.\ref{fig:one}) and chlorophyll is less sensitive to the nutrient concentrations in the presence of eddies (Fig.\ref{fig:six}). 

\subsection{Scale-propagation of a multiplicative stochastic noise}

In the last part of our analysis we investigate the impact of a stochastic Gaussian white noise on the chlorophyll dynamics across range of spatial scales. Such white noise usually represents a number of higher-complexity, scale-constrained processes that were not explicitly included into the dynamical model. If such processes have linear relationship to the dynamical model variables, their impact on the model variables will remain constrained to the (spatio-temporal) scales of those processes. However if the relationship between those processes and dynamical model variables is highly non-linear, the impact of those processes on the model variables may propagate beyond the original scale of the process. A simple example is the impact of wind stress on the vertical mixing and primary productivity in the water column: the phenomena observable on weekly time-scales, such as phytoplankton blooms (e.g. see the critical turbulence hypothesis in \cite{huisman1999critical}), may be sensitive to such details, as to whether we capture wind stress with an hourly, or 3-hourly resolution \cite{powleysensitivity}. 

\begin{widetext}

\begin{figure}
\noindent\includegraphics[width=18cm, height=7cm]{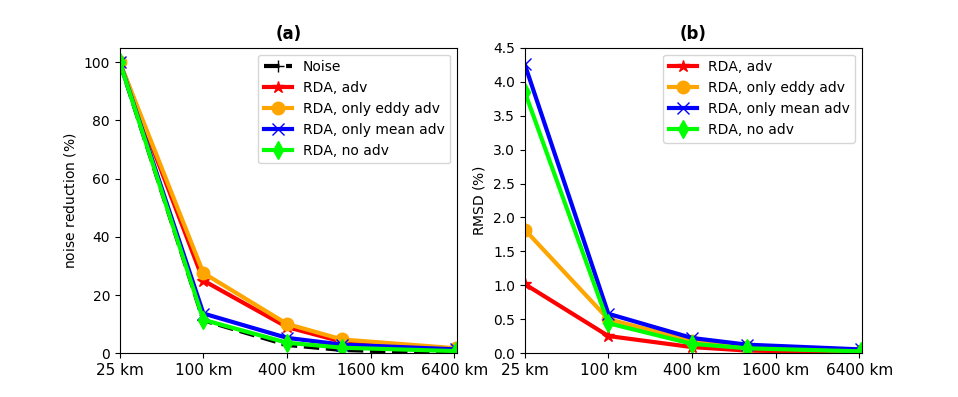}
\caption{The two panels show the impact of different drivers (e.g. mean and eddy advection, biological activity) on the propagation of a white stochastic noise in the RDA model. The panel $b$ shows the Root Mean Square Difference (RMSD) in chlorophyll between the stochastic run and the corresponding deterministic run (y-axis) vs spatial log-scale (x-axis). The RMSD values are divided by the mean 2018 chlorophyll of the deterministic run and shown in $\%$. The panel $a$ shows the same quantity, only compared (in \%) to its own value at the lowest, 25 km scale. The purpose of the panel $a$ is to show how the impact of the stochastic noise propagates through the spatial scales under different dynamical scenarios. The different scenarios are: i) the RDA model configuration with the mean and eddy currents ("RDA, adv"), ii) the RDA model with the mean currents removed ("RDA, only eddy adv"), iii) the same RDA configuration with the mesoscale eddies and diffusion removed ("RDA, only mean adv"), and iv) the RDA model without any advection ("RDA, no adv"). The panel $a$ compares the chlorophyll RMSD to the scaling of a white noise ("Noise") applied at the model resolution scale.}
\label{fig:twelve}
\end{figure}

\end{widetext}

We have run the 2017-2018 model simulation with a multiplicative white noise \cite{schenzle1979multiplicative, graham1982carleman} to account for a random variability in the growth rate parameter $P$ (Eq.\ref{E1}). The multiplicative Gaussian noise has already proven to be both realistic and useful in the population dynamics models \cite{denaro2013dynamics, valenti2016stochastic}. 
The Fig.\ref{fig:twelve} compares simulations in which the growth parameter ($P$) was perturbed by the Gaussian noise with 20\% standard deviation (corresponding to $\Delta P = \pm 1.4.10^{-8}m^{3}mmol^{-1}s^{-1}$). The random perturbations were applied as a white noise on the RDA model-grid spatio-temporal scale (25 km and 1 day).
The Fig.\ref{fig:twelve} shows the magnitude and scale-propagation of the stochastic noise impact on chlorophyll in simulations using different sets of dynamical drivers (the RDA model, the RDA model without mean currents, the RDA model without eddies, the RDA model without any advection). 
The outputs for the stochastic simulations were low-pass filtered at different scales (25 km, 100 km, 400 km, 1600 km and at the ``regional'' scale, 6400 km, where only total spatial averages were calculated) and compared with the corresponding low-pass filtered deterministic simulations (with the fixed $P$ value). The chosen metric for the comparison was the Root Mean Square Difference (RMSD). The Fig.\ref{fig:twelve}:$a$ shows the percentage of the 25 km scale RMSD that remains on scales $>25$ km. The larger is the percentage, the more is the 25 km white noise propagated to the larger scales by the model dynamics. The reduction of chlorophyll RMSD as a function of scale is compared to the scaling of the mean absolute value of the white noise originally applied to the RDA model (shown by the black dashed curve in Fig.\ref{fig:twelve}:$a$ labeled as ``Noise''). The white noise is by definition uncorrelated on the scales above 25 km, but it remains visible also on the 100-1000 km spatial scales (on the level of $<$10\%, Fig.\ref{fig:twelve}), since the low-pass filtering applied at $\geq$ 100 km scales effectively averages out the white noise over a finite number of samples, so the low-pass filtered mean will differ from the theoretical zero mean of the sampling Gaussian distribution. The number of samples increases with the spatial scale of the low-pass filter and in the limit of infinite scale the mean absolute value of the noise is precisely zero. Since the non-linear dynamics of the RDA model is expected to propagate the white noise to larger scales, the mean absolute value of the white noise applied to the RDA model is expected to reduce faster than the RMSD of chlorophyll. The black dashed curve in Fig.\ref{fig:twelve}:$a$ can be then interpreted as a ``theoretical maximum'' for the chlorophyll RMSD reduction as a function of scale, such theoretical maximum being reached when the RDA does not scale-propagate the stochastic noise. 


For a non-advective RDA model ($\vec{u}=\kappa_{T}=0$) the multiplicative noise generates at each spatial point a type of random-walk solution which is constrained to some neighborhood of the steady state solution (Eq.\ref{E1.4}). The steps of the random walk are larger in nutrient-rich areas, however this might be compensated by the fact that the convergence of a perturbed solution to the unperturbed solution might be faster in the areas with larger nutrient concentrations (Eq.\ref{E1.8}). The Fig.\ref{fig:twelve} shows that the multiplicative noise with 20\% standard deviation leads to 4\% RMSD in chlorophyll when the model has no advection, or runs with only mean advection (Fig.\ref{fig:twelve}:$b$). The RDA model without mesoscale eddies (and sub-grid eddy diffusion) does not propagate the noise to the $\geq$ 100 km scales, as the noise reduction in those simulations is close to its ``theoretical maximum'' (Fig.\ref{fig:twelve}:$b$). When mesoscale eddies (and sub-grid eddy diffusion) are included, the fluctuations in chlorophyll introduced by the stochastic noise on the 25 km scale, decrease to 2\%, or 1\% depending on whether we include also the mean currents (Fig.\ref{fig:twelve}:$b$). However, mesoscale eddies and the turbulent diffusion term introduce scale-propagation into the chlorophyll noise, with 10-30\% of the 25-km fluctuations visible on the 100-500 km scales (Fig.\ref{fig:twelve}:$a$). The reason for this scale-propagation of the chlorophyll noise is the eddy mixing, which smooths the chlorophyll noise, lowering the size of the chlorophyll random fluctuations (see the lower RMSD at the 25 km scale, Fig.\ref{fig:twelve}:$b$), but introducing larger-scale correlations to the random fluctuations. These larger scale correlations explain why the RMSD reduces comparably slowly as a function of scale (Fig.\ref{fig:twelve}:$a$).  

\subsection{Summary}

Low complexity reaction-diffusion, or reaction-diffusion-advection models have been often used to study conceptual questions in population biology, such as the critical patch size for population survival \cite{ryabov2008population}. However, a realistic simulation of phytoplankton dynamics in a specific global region is typically assumed to require a medium, or high complexity models. Here we demonstrate that for very specific purposes in a suitably tailored choice of region (e.g. tropical Pacific), the RDA model forced by a higher-complexity model outputs for nutrients and surface currents, provides a sufficiently realistic simulation for the phytoplankton chlorophyll concentrations (a proxy for primary productivity and phytoplankton biomass). The advantage of the RDA model is that the model depends only on few external inputs and model parameters, all of which are straightforward to interpret and modify. Since the model is computationally cheap to run and can be easily perturbed with a stochastic noise, one can produce almost arbitrary number of both deterministic and stochastic simulations. 

We use the RDA model to develop a multi-scale view of a driver (eddy and mean advection, eddy diffusion) impact on the chlorophyll distributions. The impact of different drivers on chlorophyll is explored in a series of simulations, where we remove specific set of drivers and analyse the changes to the chlorophyll variability on a range of spatial (25-2500 km) and temporal (1 day - 1 year) scales. 
We show that for the $1/4^{\circ}$ model, advection has a major impact on the mean chlorophyll concentrations. Turbulent diffusion has a negligible impact on the mean chlorophyll concentrations, but it is propagated by the larger scale currents and influences chlorophyll variability on a wide range of spatial and temporal scales. The scale-impact of drivers on the phytoplankton was evaluated through the magnitude of spatial, or temporal variability corresponding to the first statistical moment of chlorophyll increments (e.g. \cite{schertzer1987physical}), in the future this analysis might be extended to a more complete statistical view of chlorophyll scaling that would include also higher statistical moments.

We analysed the impact of surface nutrient decline and changes to the mesoscale eddy kinetic energy (EKE) on the mean surface chlorophyll concentrations (some changes to nutrients and EKE are projected in the future climate scenarios). The RDA model indicates that unless EKE radically changes from its current levels, chlorophyll tends to scale sub-linearly with nutrients, which implies that the chlorophyll concentrations are relatively stable with respect to the nutrient decline. However, the RDA model also shows that the chlorophyll sensitivity to nutrients goes through a sudden transition and becomes substantially larger if we minimise the EKE to 0-15\% from its current value. In the limit of vanishing EKE, chlorophyll scales with nutrients approximately linearly. We also investigate the spatial scale-propagation of a white multiplicative Gaussian noise, introduced into the RDA model on the model resolution scale. We demonstrate that the impact of the stochastic noise on the chlorophyll concentrations propagates to 100-500 km spatial scales through the mixing by eddy advection and diffusion term.


This study aims to provide an inspiration for researchers to further explore specific contexts in which low-complexity models could serve as a sufficiently realistic tool to address questions that would often be beyond the computational affordability of the higher-complexity models. The limitations of the low-complexity model need to be always recognized, but this should not mean that low-complexity models have to be always discarded as a tool of realistic modelling. Eventually the future modelling could become a multi-complexity effort, where high and medium complexity models become integrated with low-complexity models, each serving its optimal purpose while mutually achieving the desired goal with a reduced computational cost. Moreover, the low-complexity models such as the RDA model used in this study, could provide a priceless public educational tool to enhance the understanding of marine biogeochemistry in different realistic situations.   



~

{\bf Acknowledgments:} The data used to force the RDA model were obtained from Copernicus Marine Environment Service (CMEMS) and can be freely downloaded from https://resources.marine.copernicus.eu/.

\bibliography{KF_Skakala.bib}

\end{document}